\documentclass[%
 reprint,
 amsmath,amssymb,
 aps,
floatfix,
]{revtex4-2}

\usepackage{graphicx, caption, subcaption}
\usepackage{dcolumn}
\usepackage{bm}

\usepackage{siunitx}

\usepackage{hyperref}
\hypersetup{
    colorlinks=true
}

\begin{document}

\title{Simulation of in-ice cosmic ray air shower induced particle cascades}

\author{S. De Kockere}
 \email{simon.de.kockere@vub.be}
 \affiliation{Vrije Universiteit Brussel, Dienst ELEM, IIHE, Pleinlaan 2, 1050 Brussels, Belgium}
\author{K.D. de Vries}
 \email{krijn.de.vries@vub.be}
 \affiliation{Vrije Universiteit Brussel, Dienst ELEM, IIHE, Pleinlaan 2, 1050 Brussels, Belgium}
\author{N. van Eijndhoven}
 \email{nick.van.eijndhoven@vub.be}
 \affiliation{Vrije Universiteit Brussel, Dienst ELEM, IIHE, Pleinlaan 2, 1050 Brussels, Belgium}
\author{U.A. Latif}
\email{uzair.abdul.latif@vub.be}
 \affiliation{Vrije Universiteit Brussel, Dienst ELEM, IIHE, Pleinlaan 2, 1050 Brussels, Belgium}

\date{\today}

\begin{abstract}
We present simulations of the development of high-energy cosmic-ray air showers penetrating high-altitude ice layers that can be found at the polar regions. We use a combination of the CORSIKA Monte Carlo code and the Geant4 simulation toolkit, and focus on the particle cascade that develops in the ice to describe its most prominent features. We discuss the impact of the ice layer on the total number of particles in function of depth of the air shower, and we give a general parameterization of the charge distribution in the cascade front in function of $X_{max}$ of the cosmic ray air shower, which can be used for analytical and semi-analytical calculations of the expected Askaryan radio emission of the in-ice particle cascade. We show that the core of the cosmic ray air shower dominates during the propagation in ice, therefore creating an in-ice particle cascade strongly resembling a neutrino-induced particle cascade. Finally, we present the results of simulations of the Askaryan radio emission of the in-ice particle cascade, showing that the emission is dominated by the shower core, and discuss the feasibility of detecting the plasma created by the particle cascade in the ice using RADAR echo techniques.
\end{abstract}

\maketitle


\section{\label{sec:introduction}Introduction}

By proving the existence of high-energy cosmic neutrinos in the TeV-PeV energy range, the IceCube neutrino observatory opened a new and exciting window on the Universe~\cite{icecube2013}. As neutrinos can only interact weakly, they are not influenced by magnetic fields and point straight back to their source, making them the perfect candidates to perform particle astronomy. However at energies exceeding several \si{\peta\eV}, even the cubic kilometer detection volume of IceCube proves to be insufficient for collecting a significant amount of events. 

Given the long attenuation length of radio signals, the detection modules of radio neutrino observatories, typically being separated by distances of the order of \SI{1}{\km}, can efficiently cover much larger detection volumes. As such, looking for interacting neutrinos using radio detection techniques could offer a valuable extension towards these extremely high energies. 

Detecting neutrino interactions using radio signals can be performed both in a direct and an indirect way. At high neutrino energies, the build-up of a net negative charge inside a neutrino-induced particle cascade in a medium will lead to the emission of coherent radiation in the radio band. This so-called Askaryan emission offers the possibility to detect the neutrino through the direct radio detection channel~\cite{askaryan1962, ZHS1992, Alvarez1997}. Alternatively, while the cascade propagates through the medium it leaves an ionization trail, providing the possibility for indirect detection based on RADAR techniques~\cite{deVries2015, Prohira2019, Prohira2020}.

In contrast to optical Cherenkov detectors, which require very clear ice where background light is not permitted, the detection modules of in-ice radio neutrino observatories can be located in the upper layers of polar ice, situated at depths typically ranging from the surface down to \SI{200}{\m} in the ice. This significantly reduces construction costs, yet this also means that these observatories are not by construction shielded from radio emission created above the ice surface. Inevitably, anthropogenic radio signals will be picked up by such in-ice radio observatories and will need to be dealt with during data analysis. Although not their primary goal, also the Askaryan and geomagnetic radio emission from cosmic-ray-induced air showers should be visible for radio detectors located close to the surface of the ice sheet, providing a signal very similar to the one expected from an in-ice neutrino-induced particle cascade~\cite{Schroder2017, Huege2017}. If not treated properly, such signals pose an important background in neutrino searches. However, in case of a detailed understanding, detecting air shower signals could serve as a proof of principle of newly constructed neutrino observatories and could also be used to calibrate the detector modules.

Therefore, a dedicated study is required to investigate the radio emission from air showers as seen by in-ice radio detectors. By comparing simulations with observations of surface radio arrays, we know that the in-air radio emission of air showers is well understood~\cite{Nelles2014, Apel2021, Bechtol2022}. It is however a non-trivial task to extend these simulations to include in-ice radio arrays, a process where both the signals as well as the cascade move between different media.

While the in-air propagation of a cosmic-ray air shower and its radio emission has been studied in detail, only a few studies exist that investigate its propagation into ice~\cite{Razzaque2002, Razzaque2002_add, Seckel2008, Javaid2012, deVries2016, DeKockere2021} or other media~\cite{Tueros2010}. It is the in-ice propagation of the cascade which is the focus of this work. We first illustrate the general properties of high-energy cosmic ray air showers at high altitudes typical for radio neutrino detection sites. Next we describe the simulation setup that was used to simulate the propagation of cosmic ray air shower cores through high-altitude ice layers, and summarize the simulation results. We will discuss the energy deposited in the ice, the longitudinal profile of the in-ice particle shower and the lateral charge distribution in the ice. We will show that both the longitudinal profile and the lateral charge distribution in the ice can be parameterized in a universal fashion. Finally, we briefly discuss the application of our simulations to radio neutrino observatories located at high-altitude ice-sheets. In this work we show that indeed the cosmic-ray shower core propagating from air into ice should be detectable by currently existing and future Askaryan radio detectors~\cite{ANITA2009, ARA2012, ARIANNA2015, RNOG2021, PUEO2021}. Furthermore, we show that the ionization trail left in the wake of the cascade should have properties favorable for radar detection~\cite{RETCR2021}.

\section{Cosmic-ray air showers}
High-energy cosmic-rays impacting Earth's atmosphere will initiate a particle cascade, which at its maximum contains millions to billions of particles, mainly gamma rays, electrons and positrons, but also a relatively large amount of muons and hadrons is observed. Due to its statistical nature, typical global quantities are used to obtain information on the primary cascade inducing particle and its properties. For example, the total number of particles at a certain stage in the shower development, called the longitudinal profile of the shower, typically scales with the primary particle energy, and the location of the shower when it contains a maximum number of particles can be linked to the composition of the primary cascade inducing particle~\cite{Kampert2012}. More detailed parameters to use are found in the dimensions of the cascade front and the particle energy distributions. The electromagnetic component typically has a lateral extent of O(\SI{100}{\m}), but its muonic component can  reach  out to kilometer scales. The longitudinal particle spread in the cascade front is rather small and found to be ten(s) of centimeters close to the cascade axis up to several meters further out. 

To investigate the in-ice propagation of a high-energy cosmic-ray-induced air shower, we will focus on the macroscopic cascade properties. For illustration, we will use an air shower initiated by a \SI{e17}{\eV} proton impacting Earth's atmosphere with a zenith angle $\theta = 0^{\circ}$ simulated using the \mbox{\emph{CORSIKA 7.7100}} Monte Carlo code~\cite{Heck1998} and describe its general features at an observation level of \SI{2.4}{\km} above sea level, which corresponds to typical altitudes of the ice sheets in polar regions. The total number of electrons and positrons of the shower reaches its maximum at a shower depth $X_{max} = \SI{680}{g/cm^2}$, where shower depth is defined by the density of the medium integrated along the path of the shower. This is a good average value for a  \SI{e17}{\eV} proton~\cite{Buitink2016} induced air shower. More details about the simulation setup can be found in Section~\ref{sec:Simulation_setup}.

\begin{figure*}
    \includegraphics[trim={5.6cm 2cm 5.6cm 2.5cm},clip]{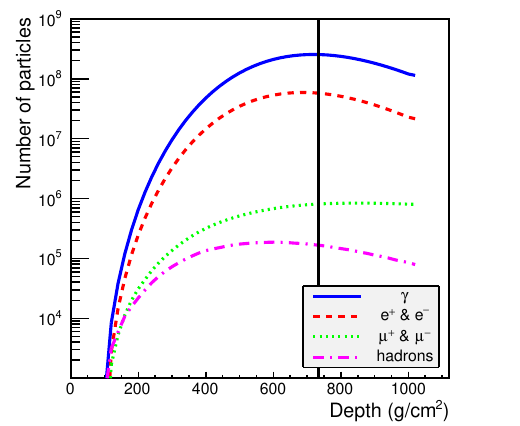}
    \includegraphics[trim={5.6cm 2cm 5.6cm 2.5cm},clip]{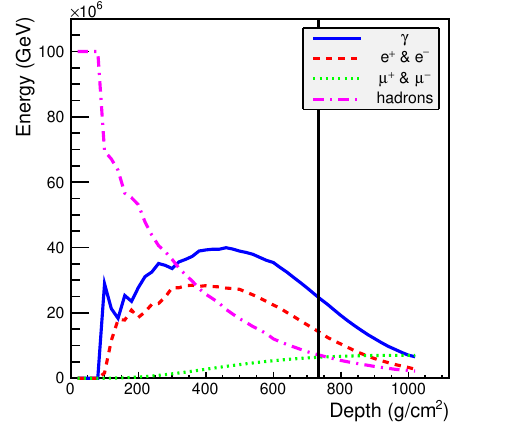}
    \caption{\label{fig:longit_profile_part} The number of particles (\emph{left}) and the distribution of the energy (\emph{right}) in an air shower initiated by a \SI{e17}{\eV} proton with zenith angle $\theta = 0^{\circ}$ reaching a given depth, simulated using \mbox{\emph{CORSIKA 7.7100}}. The black vertical line indicates an altitude of \SI{2.4}{\km}, which corresponds to a depth of \SI{734}{\g/\cm\squared}.}
\end{figure*}

Figure~\ref{fig:longit_profile_part} shows the longitudinal profile of the cascade. Here it can be seen that at \SI{2.4}{\km} altitude the high-energy particle cascade hits the surface close to its shower maximum. Looking at the number of particles we find that at this point the shower is heavily dominated by gamma rays, electrons and positrons. However, we see that the muonic and hadronic components still carry a significant fraction of the total shower energy. The muons will penetrate the ice sheet, while the hadrons will interact with the medium, feeding the electromagnetic part of the particle shower through $\pi_0$ decay. We do not expect these two components to play a major role in the development of the in-ice cascade, but for completeness we will include them in the simulations described below.

\begin{figure}
    \centering
    \includegraphics[trim={5.6cm 2cm 5.6cm 2.5cm},clip]{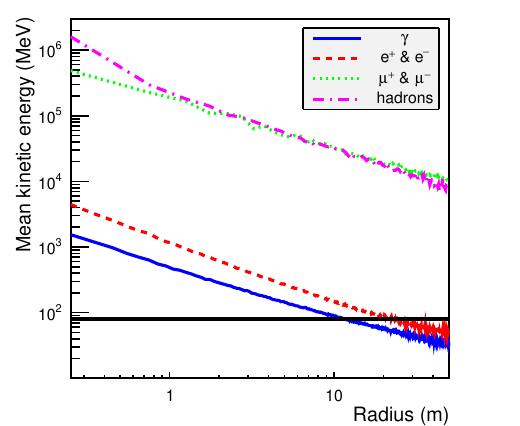}
    \caption{\label{fig:mean_kinE_dr} The average kinetic energy per particle for different components of the simulated air shower at an altitude of \SI{2.4}{\km}, in function of radius. For the electromagnetic part ($\gamma$, $e^{+}$, $e^{-}$) the average was calculated over radial bins with a bin width $\Delta r$~=~\SI{0.1}{\m}. For the muonic and hadronic part, $\Delta r$~=~\SI{0.5}{\m} was used. The black horizontal line indicates the value of \SI{80}{\MeV}.}
\end{figure}

A more detailed look into the energy distribution of the main components of the cascade is shown in Figure~\ref{fig:mean_kinE_dr}. Here we show the average energy per particle in radial bins of the order of $\Delta r = \SI{0.1}{\m}$ at an an altitude of \SI{2.4}{\km}. The horizontal line at \SI{80}{\MeV} indicates the critical energy above which electromagnetic particles are expected to shower, meaning that particles with energies above the critical energy on average will be able to interact, after which pair-production will take place, while at energies below this threshold the ionization losses become dominant and the cascade dies out. We see that electrons, positrons and gamma rays with energies above the critical energy are located close to the shower axis. The shower particles located at meter distances and more from the shower axis will vanish within a single radiation length after the cascade enters the ice. The development of the in-ice particle cascade will essentially be determined by the high-energy particles located within ten(s) of cm from the core. Interestingly, the in-ice propagation of the cosmic-ray-induced cascade can therefore be expected to perfectly mimic the tail of a high-energy neutrino-induced particle cascade in ice.

\begin{figure}
    \includegraphics[trim={5.6cm 2cm 5.6cm 2.5cm},clip]{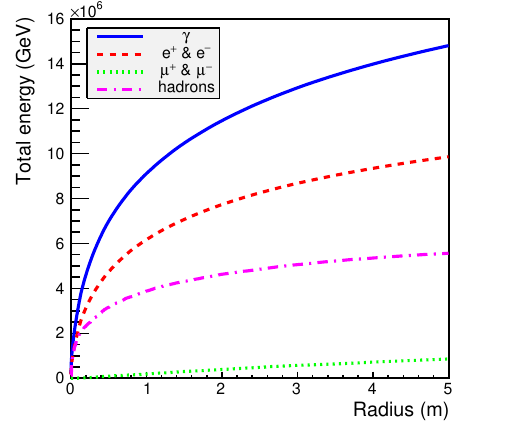}    \caption{\label{fig:cum_energy_dist} The total energy within a given radius for different components of the simulated air shower at an altitude of \SI{2.4}{\km}.}
\end{figure}

\begin{figure}
    \includegraphics[trim={5.6cm 2cm 5.6cm 2.7cm},clip]{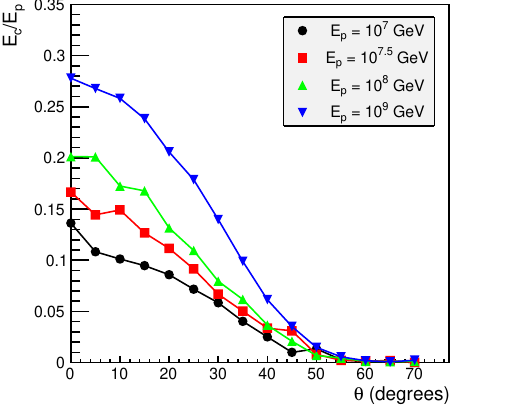}
    \caption{\label{fig:energy_cores} The energy $E_c$ contained within a radius of \SI{100}{\cm} at an altitude of \SI{2.4}{\km} in function of the energy of the primary proton $E_p$ and zenith angle $\theta$. Each point represents the average of 10 \mbox{\emph{CORSIKA}} air shower simulations.}
\end{figure}

Figure~\ref{fig:cum_energy_dist} shows the corresponding radial energy distribution of the simulated shower. The air shower clearly contains a very energy-dense core, with roughly 15\% of the primary particle's energy contained within a radius of \SI{100}{\cm}. A more detailed study of this energy-dense shower core is summarized in Figure~\ref{fig:energy_cores}. Here it is shown that the energy within the core ($<$~\SI{100}{\cm}) drops rapidly for lower primary particle energies and higher zenith angles. Therefore we expect significant in-ice cascade propagation for air showers initiated by high-energy primary particles ($E_p > $ \SI{e16}{\GeV}) and small zenith angles ($\theta < 40^{\circ}$). 

We conclude that the energy-dense core within $100$~cm of the cascade axis will be the main component determining the propagation of the particle cascade trough ice, while at larger radii a minor to even negligible contribution to the in-ice cascade development is expected.

\section{Simulation setup}\label{sec:Simulation_setup}

The main interest in this work lies in the in-ice propagation of the cosmic-ray-induced particle cascade and its associated radio detection properties. For this a dedicated simulation framework is developed using both the \mbox{\emph{CORSIKA}} Monte Carlo code and the \mbox{\emph{Geant4}} simulation toolkit.

The \mbox{\emph{CORSIKA 7.7100}} Monte Carlo code was used to generate a set of proton-induced cosmic ray air showers with different primary particle energies and zenith angles, providing both position and momentum of all particles in the showers at an altitude of \SI{2.4}{\km}. All the showers were generated using the QGSJETII-04 high energy hadronic interaction model, the GHEISHA 2002d low energy hadronic interaction model and a MSIS-E-90 atmospheric model for South Pole on December 31, 1997. Thinning was applied only for showers with primary energies $E_p \geq$ \SI{e17}{eV} on electromagnetic particles falling below $10^{-7} E_p$ with a thinning weight smaller than $10^{-7} E_p[\text{GeV}]$. For hadrons (without $\pi^0$'s) and muons a kinetic energy cut-off of \SI{0.3}{\GeV} was used. For electrons, photons and $\pi^0$'s a kinetic energy cut-off of \SI{0.003}{\GeV} was used.

The \mbox{\emph{Geant4 10.5}} simulation toolkit~\cite{Agostinelli2003} was used to propagate all the produced particles within a radius of \SI{5}{\m} of the point of impact on the surface through the ice. The simulated volume consists of multiple horizontal layers of pure ice. Each layer is \SI{1}{\cm} thick and has a constant density, with the density of the layers following the ice density gradient at the Antarctic Taylor Dome ice cap given by~\cite{Besson2008}
\begin{eqnarray}
\rho(z) = 0.460 + 0.468 \cdot (1 - e^{-0.02z}),
\end{eqnarray}
with $\rho$ the density in \SI{}{\g/\cm \cubed} and $z$ the depth in \SI{}{\meter}. We used the standard \emph{Geant4} electromagnetic physics as built by the \emph{G4EmStandardPhysics} constructor, the decay physics of long-lived hadrons and leptons as built by the \emph{G4DecayPhysics} constructor, and the radioactive decay physics as built by the \emph{G4RadioactiveDecayPhysics} constructor. \emph{Geant4} does not use kinetic energy cut-offs, but instead works with production cut-offs defined in units of length. This means that a secondary particle will only be included in the simulation if it will be able to travel a larger distance than its cut-off length. We used the default cut-off length for gammas, electrons, positrons and protons, which is \SI{1}{\mm}.

\section{Simulation results}

\subsection{Deposited energy}

\begin{figure}
    \includegraphics[trim={7.5cm 2cm 3cm 
    5cm},clip,width=\linewidth]{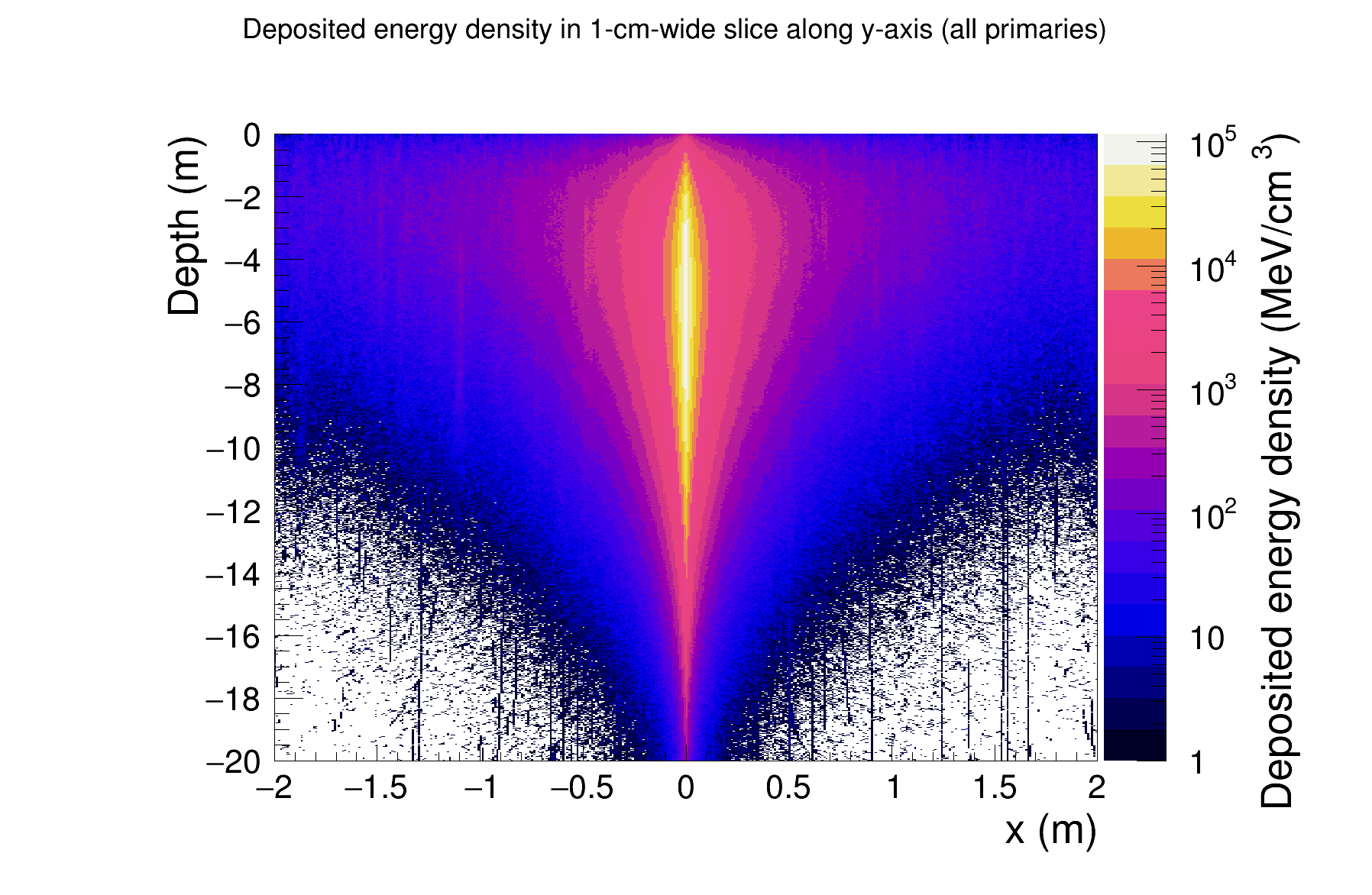}
    \caption{\label{fig:slice_histo} The energy deposited in ice by the \emph{CORSIKA} generated air shower calculated using the \emph{Geant4} module. Shown here is the deposited energy density within a vertical 1-cm wide slice going through the center of the particle shower.}
\end{figure}

\begin{figure}
    \includegraphics[trim={7.5cm 2cm 3cm 
    5cm},clip, width=\linewidth]{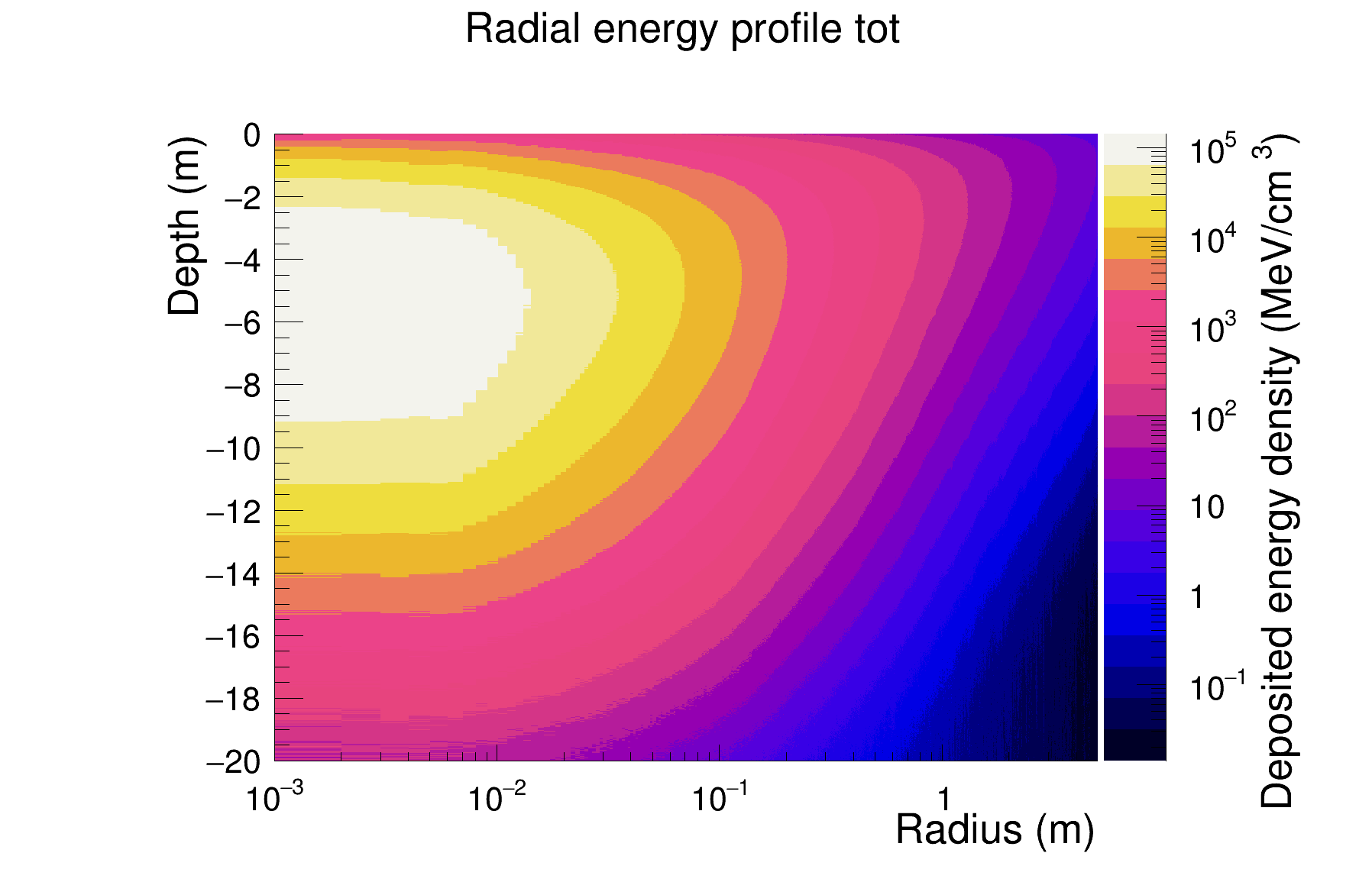}
    \caption{\label{fig:radial_histos} The energy deposited in ice by the \emph{CORSIKA} generated air shower calculated using the \emph{Geant4} module. Shown here is the radial energy density profile.}
\end{figure}

Using the \emph{Geant4} module we calculated the energy deposited by the air shower propagating through the ice. Figures \ref{fig:slice_histo} and \ref{fig:radial_histos} respectively, show the deposited energy density in the ice for a vertical 1~cm wide slice going through the center of the particle shower and the radial energy density profile for our reference shower as outlined in the previous section.

As outlined above, at meter distances the average energy of the electromagnetic component of the cascade falls well below the critical energy for showering in ice ($\sim$ 80~MeV). It follows that in this regime the cascade simply dies out over several radiation lengths equal to $X_{rad}\approx \SI{36}{\g/\cm^2}$, corresponding to a distance of $\sim 30-60\;\mathrm{cm}$ for the top layers of polar ice sheets. Toward smaller radii the average energy of the electromagnetic constituents is well above the critical energy and the cascade is observed to be showering, reaching its maximum a few meters below the ice surface.

In this work we will focus on the single reference shower presented in Figure~\ref{fig:slice_histo}. For completeness, we also show the results for different geometries in Figure~\ref{fig:slice_histos_appendix} in Appendix~\ref{sec:AppA}, where we vary the primary energy $E_p$ and zenith angle $\theta$, keeping the random seeds for the \emph{CORSIKA} shower simulations fixed and using an azimuth angle $\phi = 0^{\circ}$.

\subsection{Longitudinal profile}

\begin{figure}
    \includegraphics[trim={5.6cm 2cm 5.6cm 2.7cm},clip]{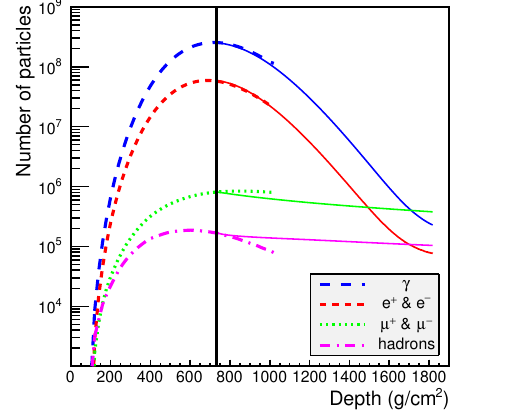}
    \caption{\label{fig:depth_plot} The number of particles in the simulated air shower in function of depth, within the full radial extend of the particle cascade. The dashed lines show the case where the particle shower propagates through air until reaching sea level. The solid lines show the number of particles with a kinetic energy above the \emph{CORSIKA} kinetic energy cut-offs given in Section~\ref{sec:Simulation_setup} when propagating the shower through ice at an altitude of \SI{2.4}{\km}, which corresponds to a depth of \SI{734}{\g/\cm\squared}, indicated by the black vertical line.} 
\end{figure}

The total number of particles as function of depth for the \emph{CORSIKA} generated air shower is shown in Figure~\ref{fig:depth_plot}. The dashed lines show the case where the shower propagates through air until reaching sea level. The solid lines show the particle shower development when the shower is propagated through ice at an altitude of \SI{2.4}{\km} using the \emph{Geant4} module, only taking into account particles with a kinetic energy above the \emph{CORSIKA} kinetic enery cut-offs stated in Section~\ref{sec:Simulation_setup}. The particle distributions are obtained over the full radial extend of the particle cascade and shown as function of depth.

From Figure~\ref{fig:depth_plot} it immediately follows that the electromagnetic part of the cascade as function of depth is hardly affected by the air-ice boundary. Therefore, parameterizations already available in literature, such as the Gaisser-Hillas parameterization~\cite{Gaisser1977}, should be well suited to describe the total particle number as function of depth for the in-ice cascade. Equivalently, one could run a full \emph{CORSIKA} simulation using our atmosphere as medium and consider the obtained depth profile valid for any given air-ice boundary. For hadrons and muons a slight deviation is observed. This is attributed to charged pion interactions, which for a more dense medium are favored over their decay. Although these interactions in principle feed the electromagnetic component by $\pi^0$ production, their number is negligible with respect to the total number of gamma's, electrons, and positrons inside the cascade. On the other hand, as the creation of muons is suppressed, the muonic component starts to decline. This behavior was also seen in the simulation described in~\cite{Tueros2010}, which propagated particle cascades underground.

\subsection{Lateral charge distribution}

\begin{figure}
    \includegraphics[trim={2.4cm 0cm 1.1cm 1.7cm},clip, width=\linewidth]{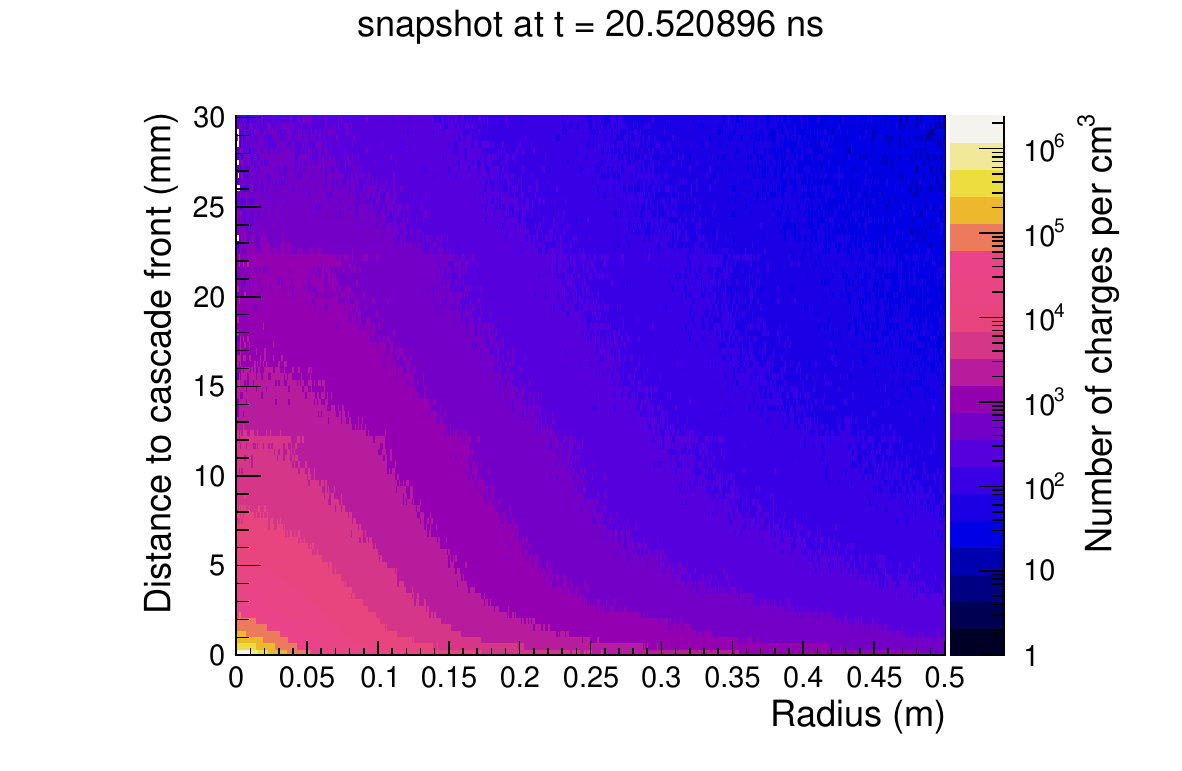}
    \caption{\label{fig:radius_depth_snapshot} A snapshot of the simulated shower, showing the radial profile of the number of charges per unit volume versus the distance along the shower axis to the cascade front. The depth of the cascade front with respect to the ice surface at this time is \SI{300}{g/cm^2}, which corresponds to a total depth including the traversed atmosphere of \SI{1034}{g/cm^2}.}
\end{figure}

The particle distributions inside the cascade front are expected to change significantly once the cascade penetrates the ice. The longitudinal extent of the cascade front itself will be very small, as illustrated in Figure~\ref{fig:radius_depth_snapshot}. For coherent processes, where the longest projected dimensions are those that are relevant, it can therefore be well approximated by a delta-distribution.

The lateral particle distribution inside the cascade is expected to be governed by multiple Coulomb scattering and hence will mainly be a function of the density of the medium. It can be described by the $w_1(r)$ distribution, with $r$ the radius in the shower axis frame. By definition $w_1(r)dr$ represents the number of charged particles within the interval $[r, r + dr]$ at a given time, normalized such that $\int_{0}^{R_0}w_1 dr = 1$. As we are mostly interested in the region close to the axis where the highest particle densities are found, we use the value $R_0 = \SI{0.2}{m}$. To calculate this distribution at a given time $t$, we construct a histogram of fixed bin width $\Delta r = \SI{1}{\mm}$ during the simulation, such that every bin represents the number of charged particles at time $t$ with a radius $r$ within the corresponding bin limits. The radius $r$ is taken with respect to the shower axis. From the histogram the $w_1(r)$ distribution can be constructed by dividing each bin value by the bin width $\Delta r$ and the total number of charges at time $t$ with $r < R_0$. Since most particles will be at the shower front, this essentially describes the cascade front particle distribution.

\begin{figure}
    \includegraphics[trim={5.6cm 2cm 5.6cm 2.7cm},clip]{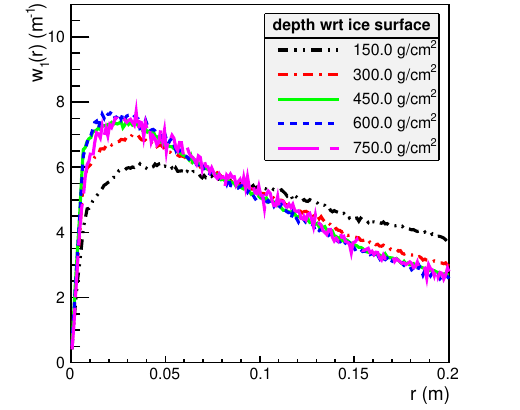}
    \caption{\label{fig:w1_radius_charges_ref_shower} The $w_1(r)$ distributions for the reference shower. Each distribution corresponds to a different depth of the cascade front, going from \SI{150}{\g/\cm\squared} (\SI{3.2}{\m}) up to \SI{750}{\g/\cm\squared} (\SI{14.4}{\m}) with respect to the ice surface, as indicated in the legend.}
\end{figure}

The $w_1(r)$ distribution for the reference shower at different time values is shown in Figure~\ref{fig:w1_radius_charges_ref_shower}. Each time value $t$ is indicated by the depth of the cascade front $X$ with respect to the ice surface, using $t = L/c_0$, with $c_0$ the velocity of light in vacuum and $L$ the distance traveled along the shower axis such that $\int_{0}^{L} \rho dl = X$. Initially an on-set effect is seen, given by a rather broad distribution toward larger radii. As outlined before, the average energy at larger radii is not large enough to produce an in-ice shower, and hence this part of the cascade quickly vanishes. Indeed, after several radiations lengths, a stable distribution is found, with a typical width following that of an in-ice neutrino-induced particle cascade.

\begin{figure}
    \includegraphics[trim={5.6cm 2cm 5.6cm 2.7cm},clip]{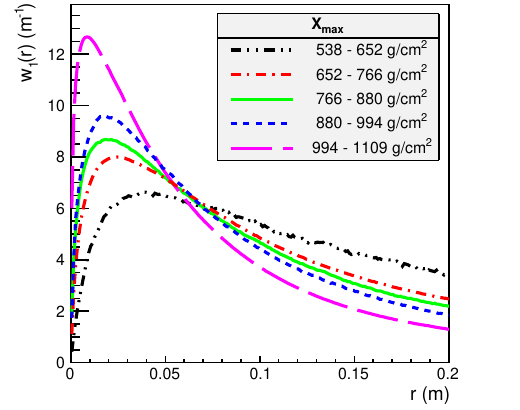}
    \caption{\label{fig:w1_radius_charges_averages} The average $w_1(r)$ distributions at a cascade front depth of \SI{450}{\g/\cm\squared} with respect to the ice surface. Indicated in the text boxes of the plots are the $X_{max}$ intervals for which the corresponding average $w_{1}$ distribution was calculated.}
\end{figure}

\begin{figure*}
    \includegraphics[trim={5.8cm 2cm 6.5cm 2cm},clip, width=0.32\textwidth]{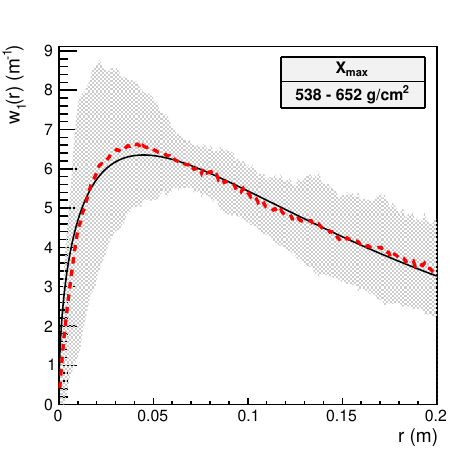}
    \includegraphics[trim={5.8cm 2cm 6.5cm 2cm},clip, width=0.32\textwidth]{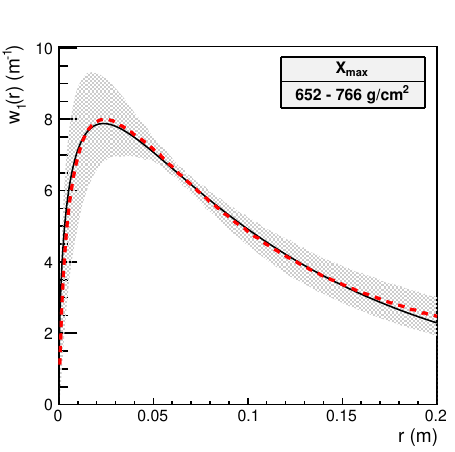}
    \includegraphics[trim={5.8cm 2cm 6.5cm 2cm},clip, width=0.32\textwidth]{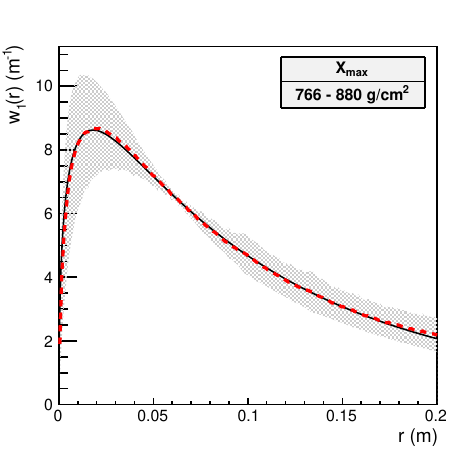}
    \includegraphics[trim={5.8cm 2cm 6.5cm 2cm},clip, width=0.32\textwidth]{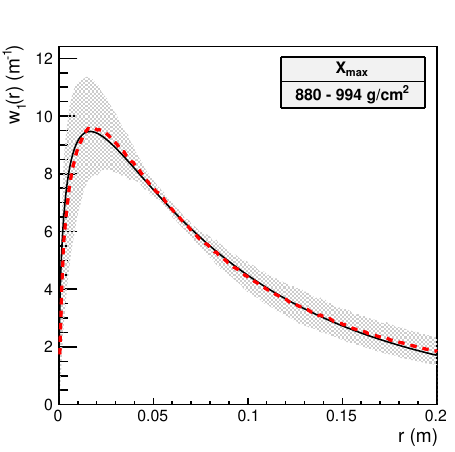}
    \includegraphics[trim={5.8cm 2cm 6.5cm 2cm},clip, width=0.32\textwidth]{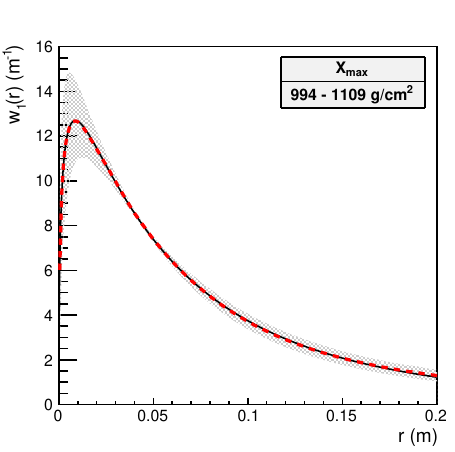}
    \caption{\label{fig:w1_radius_charges_averages_detail} The average $w_1(r)$ distributions at a cascade front depth of \SI{450}{\g/\cm\squared} with respect to the ice surface, shown by the red dashed curves. Indicated in the plot are the $X_{max}$ intervals for which the corresponding average $w_{1}$ distribution was calculated. The grey band shows the standard deviation, assuming the $w_1(r)$ distributions are normally distributed within the $X_{max}$ groups, and are interpreted as the standard deviations of the points, to take into account the fluctuation of the $X_{max}$ values of the showers within the given intervals. The solid black line shows the fit to the average $w_1(r)$ distribution following Equation~\ref{eq:fit_function}.}
\end{figure*}

To account for shower-to-shower fluctuations and the dependence on primary energy $E_p$ and zenith angle $\theta$, we have constructed 10 different simulation sets, each covering $E_p$ values of $\SI{e16}{\eV}$ to $\SI{e18}{\eV}$ in steps of half a decade and $\theta$ values of $\SI{0}{\degree}$, $\SI{15}{\degree}$ and $\SI{30}{\degree}$, resulting in 150 showers in total. As we do not expect a dependence on the azimuth angle, we keep it at $\phi = 0^{\circ}$. Within each set we used the same \emph{CORSIKA} random generator seeds for all primary energy and zenith angle combinations. In order to arrive at a very simple and straightforward parameterization of the $w_1(r)$ distributions, we investigated the dependence of the $w_1(r)$ distributions on $X_{max}$, the depth at which the shower reaches its maximum number of electrons and positrons assuming it develops in air without meeting an ice sheet. Ignoring other parameters like primary energy and zenith angle, we group the showers based on $X_{max}$, construct the $w_{1}(r)$ distributions at a given time value and calculate the average $w_{1}(r)$ distribution for each group. The average $w_1(r)$ distributions for a cascade front depth of \SI{450}{\g/\cm\squared} with respect to the ice surface are shown in Figure~\ref{fig:w1_radius_charges_averages} and Figure~\ref{fig:w1_radius_charges_averages_detail}. The figures show a clear trend between the distributions and $X_{max}$. Showers reaching their maximum number of particles early will have a less energy dense core when reaching the ice surface, resulting in a broader $w_1(r)$ distribution.

We found that the $w_1(r)$ distributions can be well described by the analytical expression

\begin{equation}\label{eq:fit_function}
    W(r) = \frac{1}{A} \sqrt{r} e^{-\left(r/b\right)^c},
\end{equation}

with the values of the fit parameters $b$ and $c$ depending on $X_{max}$ of the shower. The value of $A$ is determined by the normalization and is given by
\begin{equation*}
    A = \frac{b^{3/2}}{c} \left\{ \Gamma\left( \frac{3}{2c} \right) - \Gamma \left( \frac{3}{2c}, \Big( \frac{R_0}{b} \Big)^c \right) \right\},
\end{equation*}
with $\Gamma(x)$ the gamma function and $\Gamma(a, x)$ the upper incomplete gamma function. The fits to the average $w_1(r)$ distributions are shown in Figure~\ref{fig:w1_radius_charges_averages_detail}, with the corresponding fit parameter values listed in Tabel~\ref{tab:fit_vals} of Appendix~\ref{sec:AppB}.

\begin{figure*}
    \includegraphics[trim={5.8cm 2cm 6.5cm 2cm},clip, width=0.32\textwidth]{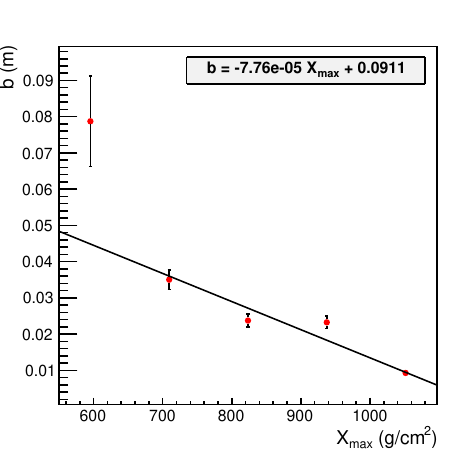}
    \includegraphics[trim={5.8cm 2cm 6.5cm 2cm},clip, width=0.32\textwidth]{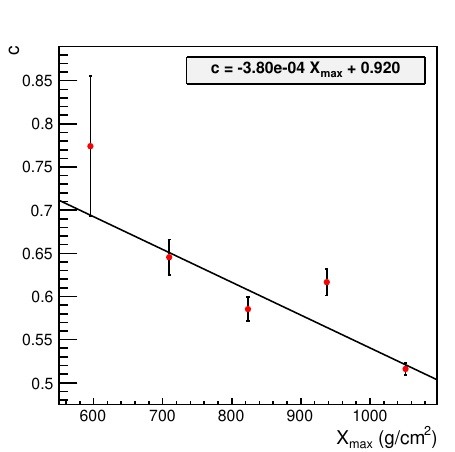}
    \caption{\label{fig:params_depth} The parameters of Equation~\ref{eq:fit_function} in function of $X_{max}$ of the shower at a cascade front depth of \SI{450}{\g/\cm\squared} with respect to the ice surface. For every interval we have taken the mean $X_{max}$ value, indicated by the red dots. The error bars are calculated by interpreting the grey bands from Figure~\ref{fig:w1_radius_charges_averages_detail} as the standard deviations of the points, so that they indicate the variation due to the fluctuation of the $X_{max}$ values of the showers within the given intervals. The black lines show the linear fits, which are given in the text boxes of the plots.}
\end{figure*}

\begin{figure}
    \includegraphics[trim={5.6cm 2cm 5.6cm 2cm},clip]{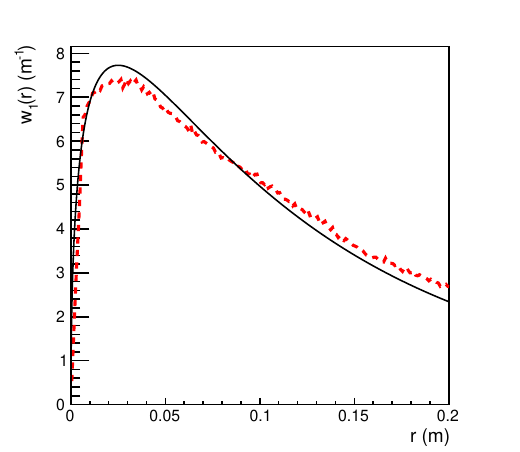}
    \caption{\label{fig:reverse_w1} The $w_1(r)$ distribution for the reference shower at a cascade front depth of \SI{450}{\g/\cm\squared} with respect to the ice surface. The red dashed line shows the distribution derived directly from the \emph{Geant4} simulation, also shown in Figure~\ref{fig:w1_radius_charges_ref_shower}. The black solid line shows the reconstructed $w_1(r)$ distribution, using the linear fits for the parameters $b$ and $c$ evaluated at $X_{max}$~=~\SI{680}{\g/\cm\squared}, combined with Equation~\ref{eq:fit_function}.}
\end{figure}

Finally, we give the fit parameters $b$ and $c$ from Equation~\ref{eq:fit_function} as a function of $X_{max}$, shown in Figure~\ref{fig:params_depth}. Here we use the mean $X_{max}$ values of the different intervals, and present linear fits which can be used to reconstruct the $w_1(r)$ distribution at a given depth in the ice for a random shower, given its $X_{max}$ value. The parameters $b$ and $c$ show a larger uncertainty for the lowest $X_{max}$ value, which is attributed to the smaller amount of particles reaching the ice surface in these type of showers, leading to larger statistical fluctuations. Figure~\ref{fig:reverse_w1} shows the reconstructed $w_1(r)$ distribution for the reference shower at a cascade front depth of \SI{450}{\g/\cm\squared} with respect to the ice surface, using the linear fits for the parameters $b$ and $c$ evaluated at $X_{max}$~=~\SI{680}{\g/\cm\squared}, combined with Equation~\ref{eq:fit_function}.

\section{Applications}

Here we discuss the feasibility of detecting the in-ice particle cascades induced by high energy cosmic ray air showers. We consider both the detection of the direct Askaryan radio emission as well as its possible detection through RADAR echo techniques.

\subsection{Askaryan radio emission}
Given that the properties of the in-ice particle cascade are very similar to a high-energy neutrino-induced cascade, we expect similar Askaryan radio emission for both cases. If not identified, this signal could pose a strong background for in-ice Askaryan detectors searching for high-energy neutrino interactions inside the dense ice sheet. On the other hand, if identified, due to its strong resemblance with the neutrino-induced cascade, this signal could provide the in-nature proof-of-principle of the Askaryan method to detect in-ice neutrino-induced particle cascades. A second application is found in its use as a possible calibration source.

To estimate the direct Askaryan radio emission of such an in-ice particle cascade we implemented the so-called end-point formalism within the \emph{Geant4} setup, using the code developed for the SLAC T-510 experiment as an example~\cite{Zilles2017, Bechtol2022}. This formalism was developed for calculating the radio emission by a charged particle following an arbitrary path through a medium. The path of the particle is divided in small segments. For each of the segments the emission is calculated as if the charge got instantly accelerated at the start-point of the segment from rest to its velocity $\vec{\beta}^*$, and instantly decelerated to a stop at the end-point of the segment. This approach fits naturally within the \emph{Geant4} framework, which provides particle trajectories in segmented steps. It can be shown that the contribution to the induced electric field received by an antenna at a certain point $\vec{x}$ at a time $t = R/(c/n)$ following this approach is given by~\cite{James2011}

\begin{eqnarray}
\vec{E}_{\pm}(\vec{x},t) = \pm \frac{1}{\Delta t} \frac{q}{c}\left( \frac{\hat{r} \times [\hat{r} \times \vec{\beta}^*]}{|1-n\vec{\beta}^* \cdot \hat{r}|R} \right),
\end{eqnarray}
with $\Delta t$ the sampling time interval of the observer, $q$ the charge of the particle, $\hat{r}$ the direction from the start/end-point towards the antenna, $R$ the distance between the start/end-point and the antenna and $n$ the index of refraction of the medium. The plus sign should be used for the instant acceleration and the minus sign for the instant deceleration. The viability of this framework has also been demonstrated by \emph{CoREAS}, an extension of \emph{CORSIKA} designed to calculate the radio emission of air showers, which is shown to agree well with experimental results~\cite{Nelles2014}. 

It should be noted, however, that the simulations presented in this work assume a constant index of refraction $n$, which might be an oversimplification when considering the upper layers of polar ice sheets. Nevertheless, as a first investigation, these results should provide a good order of magnitude indication of the expected radio signal strength. A more thorough implementation considering the detailed radio signal propagation inside the top layer of polar ice sheets will be discussed in future work.

Here we present the first results using the end-point approach. We calculate the expected signal in point-like antennas of the in-ice cascade for the reference shower, varying the primary energy and zenith angle but keeping the random seeds for the \mbox{\emph{CORSIKA}} simulations fixed and using an azimuth angle $\phi = 0^{\circ}$, as outlined before. We use a refractive index of $n=1.52$, a good average for the typical top layer of ice at polar regions. Note that the results presented here do not include radio emission created during the propagation of the particle shower through air.

\begin{figure*}
	\includegraphics[width=0.75\textwidth]{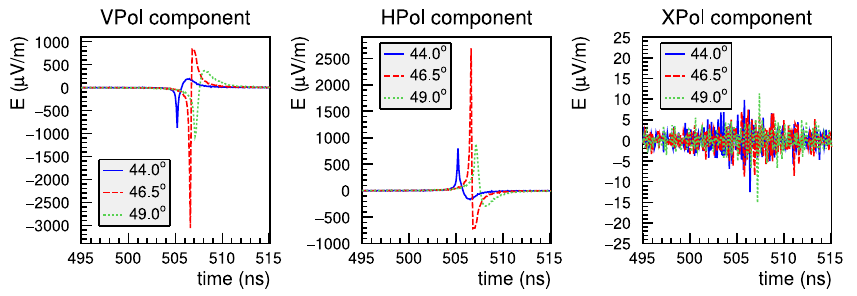}   
	\caption{\label{fig:three_traces_comb} The expected electric field components from the Askaryan radio emission of the reference shower hitting the ice surface in function of time, as seen by a point-like antenna at a distance $\SI{100}{m}$ from the point of impact of the shower core with the ice surface for different viewing angles. The viewing angle of the antenna is defined clockwise with respect to the shower axis, measured from the point of impact on the ice surface. Time $t= 0$ refers to the moment when the shower core hits the ice surface.}
\end{figure*}

Figure~\ref{fig:three_traces_comb} shows the three different components of the electric field as seen by the observer, for a primary energy $E_p = \SI{e17}{\eV}$ and zenith angle $\theta = \SI{0}{\degree}$. The antenna is placed at a distance $\SI{100}{m}$ from the point of impact of the shower core with the ice surface, in the plane corresponding to an azimuth angle $\phi = 0^{\circ}$. The viewing angle of the antenna is varied, as indicated in the figure, where it is defined clockwise with respect to the shower axis, measured from the point of impact on the ice surface.

The vertical (VPol) and horizontal (HPol) component in the $\phi = 0^{\circ}$ plane show the expected bipolar signal, reaching magnitudes well above typical detection thresholds of $O(100\;\mathrm{\mu V / m})$. The changes in the amplitude for the VPol component coincide with similar changes in amplitude of the HPol component, indicating radial polarization. As the particles in the cascade are not perfectly contained in the $\phi = 0^{\circ}$ plane, the component perpendicular to this plane (XPol) is non-zero, but significantly smaller than the VPol and HPol components.

The value for the index of refraction $n = 1.52$ corresponds to a Cherenkov angle of $49.0^{\circ}$. However, we see that the amplitude of the observed electrical field is highest around a smaller viewing angle of $46.5^{\circ}$. The viewing angle is measured from the point of impact on the ice surface, while most of the radiation is emitted deeper into the ice. Therefor, this shift towards a smaller viewing angle is to be expected. A viewing angle of $46.5^{\circ}$ measured from the ice surface gives a viewing angle of $49.0^{\circ}$ measured from the point $\SI{5.8}{m}$ down in the ice, which corresponds well with the depth of maximum energy deposit of the shower core shown in Fig~\ref{fig:slice_histo}~and~\ref{fig:radial_histos}. This also agrees with the different arrival times of the signals in the antennas.

\begin{figure*}
	\includegraphics[width=0.8\textwidth]{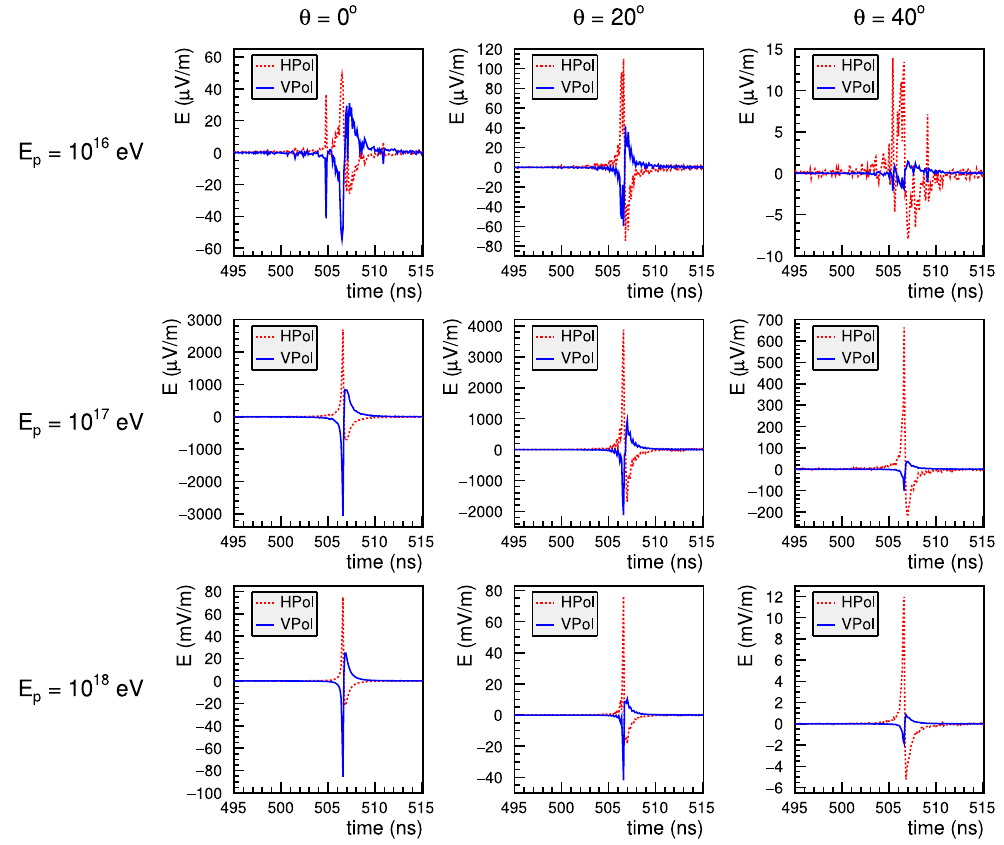}    
	\caption{\label{fig:antenna54_comb} The expected Hpol and VPol components from the Askaryan radio emission of the reference shower hitting the ice surface in function of time, as seen by a point-like antenna at a distance $\SI{100}{m}$ from the point of impact of the shower core with the ice surface. The viewing angle of the antenna is $46.5^{\circ}$ with respect to the shower axis, measured from the point of impact on the ice surface. Time $t= 0$ refers to the moment when the shower core hits the ice surface. We vary the primary energy $E_p$ and zenith angle $\theta$ for the reference shower. Different rows indicate different primary energies, while different columns indicate different zenith angles, as indicated in the figure.}
\end{figure*}

Figure~\ref{fig:antenna54_comb} shows the HPol and VPol components of the antenna at a viewing angle of $46.5^{\circ}$ measured from the point of impact on the surface, now varying the primary energy $E_p$ and zenith angle $\theta$ of the shower. In general, a higher value for the primary energy will result in a higher amplitude for the electric field, since more energy will reach the ice surface. A higher value for the zenith angle means the air shower will travel longer distances through air, causing less energy to reach the surface and resulting in a lower amplitude for the electric field. At a zenith angle of $40^{\circ}$ the drop in amplitude is clearly shown in the figure. However, we see that for this particular example, for primary energies $E_p = \SI{e16}{eV}$ and $E_p = \SI{e17}{eV}$ the amplitude of the electric field at zenith angle $\theta = 20^{\circ}$ is higher compared to their $\theta = 0^{\circ}$ counterparts. This can be attributed to statistical fluctuations in the shower development process in the air, leading to slightly higher values for the deposited energy density in the ice at $\theta = 20^{\circ}$, as shown in Fig~\ref{fig:slice_histos_appendix} in Appendix~\ref{sec:AppA}. Fixing the random seeds for the \emph{CORSIKA} simulation only fixes the properties of the first interactions, while changing the total travel distance in air introduces statistical fluctuations further down in the particle cascade development.

As the electric field is radially polarized, the zenith angle of the primary particle also influences the relative contributions of the HPol and VPol components. At a viewing angle of $46.5^{\circ}$ and a zenith angle $\theta = 0^{\circ}$, the magnitude of the HPol and VPol component are expected to be very similar. At higher zenith angle values, the ratio of the HPol component to that of the VPol component will increase. This behavior is clearly visible in Figure~\ref{fig:antenna54_comb}.

\begin{figure}
    \includegraphics[trim={5.4cm 2cm 6.5cm 2.5cm},clip]{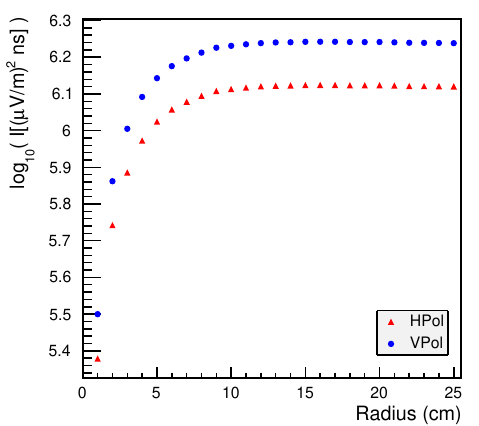}
    \caption{The integrated intensity at a viewing angle of $46.5^{\circ}$ measured from the point of impact on the surface, in function of the outer radius of the reduced particle footprint of the reference air shower (using $E_p = \SI{e17}{\eV}$ and $\theta = 0^{\circ}$).}
    \label{fig:int_intensity}
\end{figure}

To illustrate once more the importance of the shower core, we limit the particle footprint of the reference air shower to all particles within a radius $r = \SI{1}{cm}$ from the shower core, using $E_p = \SI{e17}{\eV}$ and $\theta = 0^{\circ}$, and simulate its contribution to the electric field at a viewing angle of $46.5^{\circ}$ with respect to the point of impact on the surface. Next we calculate the integrated intensity at the antenna position, here defined as $I = \int E(t)^2 dt$, where the integral is taken over a \SI{400}{\ns} time interval starting \SI{20}{\ns} before the first non-zero value of $E(t)$. As such we find a measure for the intensity of the emission of the in-ice shower, using only a reduced footprint of the air shower. We gradually increase the radius of the reduced footprint by dividing the full footprint of the air shower on the ice surface into concentric rings of width $\Delta r = \SI{1}{cm}$. For each ring we simulate the contribution to the electric field at the antenna position, add it to the total electrical field and calculate the integrated intensity. The result is shown in Figure~\ref{fig:int_intensity}. We see that the integrated intensity reaches a maximum value around a radius of $\SI{10}{cm}$, for both the HPol and VPol component, showing that the particles within this radius dominate the radio emission.

\subsection{RADAR technique}
It was recently shown at the Stanford Linear Accelerator Center (SLAC) that high-energy particle cascades in dense media can be probed using the radar echo technique~\cite{Prohira2020}. This indicates that this method could be used to detect neutrino-induced particle cascades in ice. The in-situ proof-of-principle in nature is however still to be shown. We have already demonstrated that cosmic ray air showers create particle cascades in ice with properties very similar to neutrino-induced cascades. Therefore, the radar detection of these in-ice cosmic-ray-induced particle cascades would show the proof-of-principle of the method in nature.

To investigate the effectiveness of the detection of an in-ice particle cascade through the reflection of radio waves on the plasma created in the ice, we study the behaviour of the corresponding plasma frequency $\omega_p$. This parameter gives us an estimation of the reflection capability of the plasma. The plasma frequency can be related to the free charge density $n_q$ of the plasma using~\cite{deVries2015}

\begin{eqnarray}
\omega_p = 8980 \sqrt{n_q[\text{cm}^{-3}]} \text{ Hz}.
\end{eqnarray}

Assuming the bulk of the free charges in the plasma are electrons, we can estimate the free charge density from the deposited energy density $\rho_E$ in the ice using \SI{50}{eV} as a typical ionization energy. This gives us

\begin{eqnarray}
n_q = \rho_E/(\SI{50}{\eV}).
\end{eqnarray}

\begin{figure}[h!]
    \includegraphics[trim={7.5cm 2cm 3cm 
    5cm},clip,width=\linewidth]{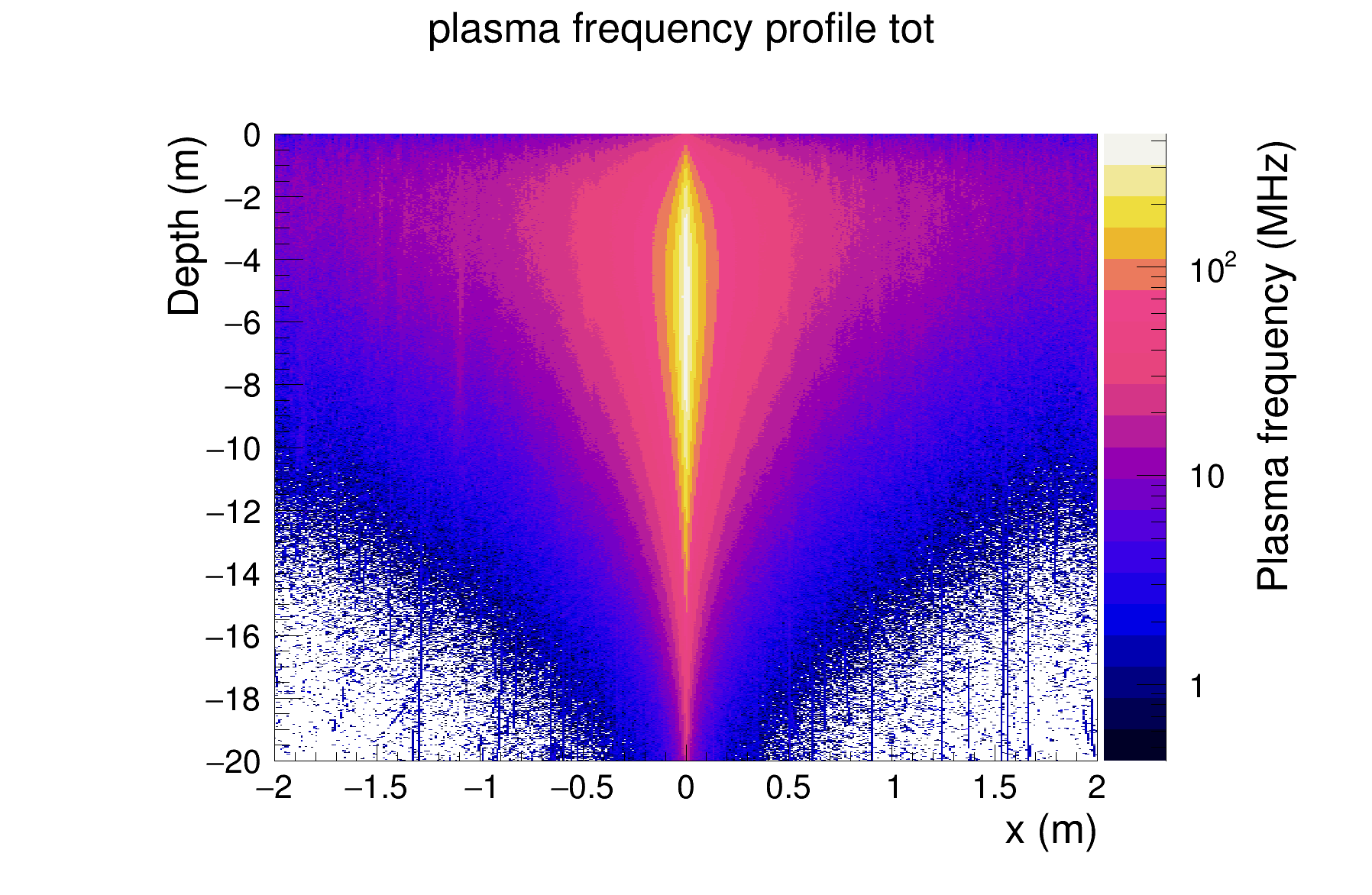}
    \caption{\label{fig:slice_histo_freq} The estimated plasma frequency $\omega_p$ of the plasma created in ice by the reference shower.}
\end{figure}

Figure~\ref{fig:slice_histo_freq} shows the plasma frequency corresponding to the deposited energy density in ice shown in Figure~\ref{fig:slice_histo}. As can be seen the energy dense core reaches plasma frequencies of \SI{100}{MHz} and above, very similar to the plasma frequencies found at the T-576 experiment at SLAC, and as shown in~\cite{RETCR2021} it should be able to efficiently reflect radio emission.

\section{Conclusions}

Using a combination of the \emph{CORSIKA 7.7100} Monte Carlo code and the \emph{Geant4 10.5} simulation toolkit, we simulated the propagation of high-energy cosmic ray air showers through a realistic ice layer at an altitude of \SI{2.4}{\km}. The primary goal is to understand its properties important for studying such events using radio detection techniques.

First we discussed the general properties of high-energy cosmic ray air showers at such a high altitude. We found that they are in general close to shower max. Particles close to the shower axis ($< \SI{100}{\cm}$) will still contribute to the development of the particle shower, while further out the shower is dying out. We see that the shower contains a very energy-dense core, which will be the main component determining the development of the cascade when propagating through ice. It can therefore be expected that it will have properties very similar to neutrino-induced particle cascades in ice.

Next we studied the general properties of the in-ice particle cascade, created by the propagation of a high-energy cosmic ray air shower through the ice layer. We calculated the deposited energy density in the ice, confirming that the shower core is still developing indeed, reaching a maximum at a few meters below the ice surface. We showed that the longitudinal profile of the electromagnetic part of the cascade is not affected by the air-ice boundary when expressed in function of depth $X$, meaning existing parameterizations can be used to describe this profile. We described the lateral charge distribution in function of radius and depth of the shower front with respect to the ice surface, and concluded that after several radiation lengths a stable distribution is found with a similar typical width of an in-ice neutrino-induced particle cascade. Furthermore, we found a clear trend between the exact form of the lateral distribution and the $X_{max}$ value of the cosmic ray air shower. We showed that it is possible to describe the lateral distribution at a given depth of the shower front using an analytical expression, of which the free parameters are determined by the value of $X_{max}$. This can be used for analytical and semi-analytical calculations of the expected Askaryan radio emission of the in-ice particle cascade~\cite{Alvarez2009, Hanson2017, Alvarez2010, Glaser2020}.

Finally we discussed the feasibility of detecting the Askaryan radio emission created by these in-ice particle cascades, as well as the possibility of reflecting emitted radio waves off of the plasma created by the in-ice particle cascades. We used the end-point formalism to simulate the Askaryan radio emission, and found that for a given antenna position the electric field reaches a magnitude well above typical detection thresholds of $O(10-100\;\mathrm{\mu V / m})$. Furthermore, we concluded that the energy-dense core of the particle shower dominates the radio emission. From the deposited energy density in the ice we estimated the free charge density of the corresponding plasma, and found that close to the shower axis the plasma frequency reaches values required for a realistic RADAR setup to detect these in-ice cascades.

\begin{acknowledgments}
This work was supported by the Flemish Foundation for Scientific Research (FWO-G007519N) and the European Research Council under the European Unions Horizon 2020 research and innovation program (No 805486 - K.~D. de Vries). We would like to thank D. Seckel, C. Glaser and S. Barwick for ongoing discussions about the presented work, as well as S. Prohira and T. Huege for their valuable feedback.
\end{acknowledgments}

\nocite{*}

\bibliography{references}
\clearpage 

\onecolumngrid
\appendix
\section{Deposited energy densities}\label{sec:AppA}

\begin{figure*}[h!]
	\centering
	\begin{subfigure}{\textwidth}
		\centering
		\includegraphics[width=\linewidth]{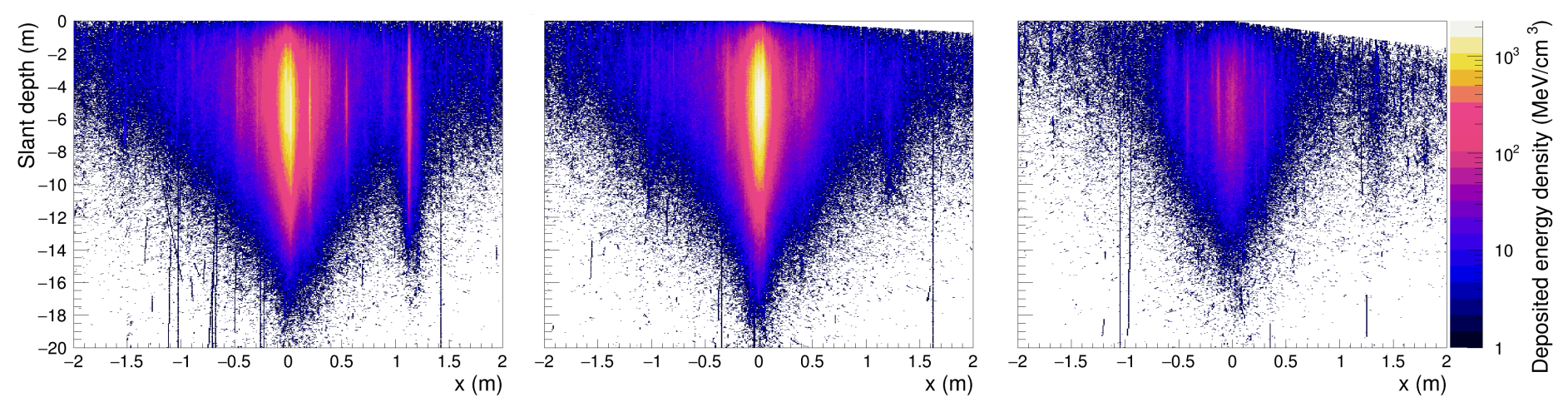}
		\caption{Deposited energy densities for $E_p = \SI{e16}{eV}$. From left to right, the zenith angle values are $\theta = 0^{\circ}$, $\theta = 20^{\circ}$ and $\theta = 40^{\circ}$. }
		\label{fig:slice_histos_1e7} 
	\end{subfigure}
	\hfill
	\begin{subfigure}{\textwidth}
		\centering
		\includegraphics[width=\linewidth]{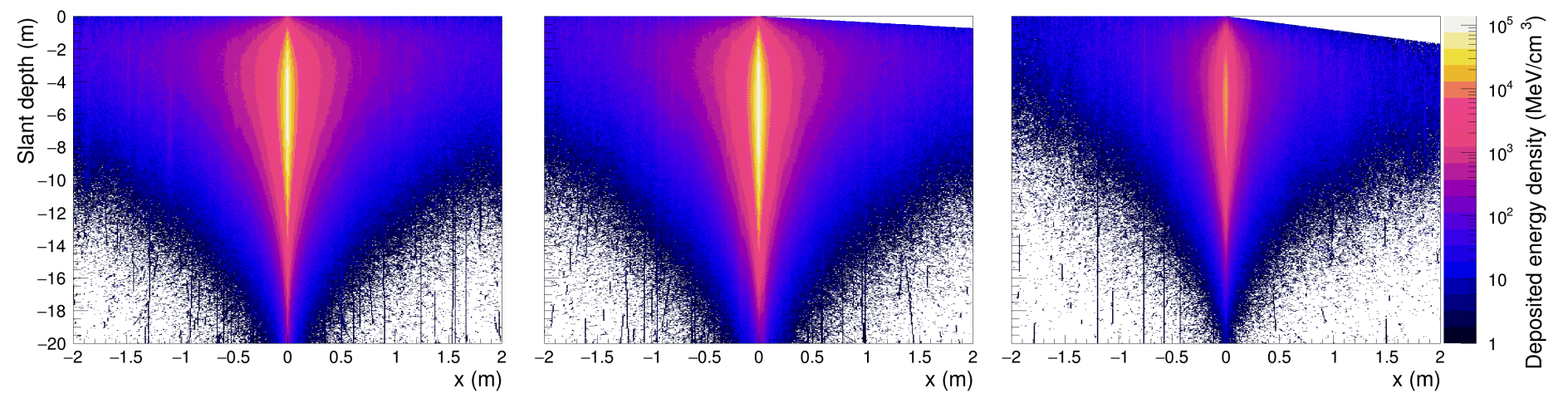}
		\caption{Deposited energy densities for $E_p = \SI{e17}{eV}$. From left to right, the zenith angle values are $\theta = 0^{\circ}$, $\theta = 20^{\circ}$ and $\theta = 40^{\circ}$. }
		\label{fig:slice_histos_1e8} 
	\end{subfigure}
	\hfill
	\begin{subfigure}{\textwidth}
		\centering
		\includegraphics[width=\linewidth]{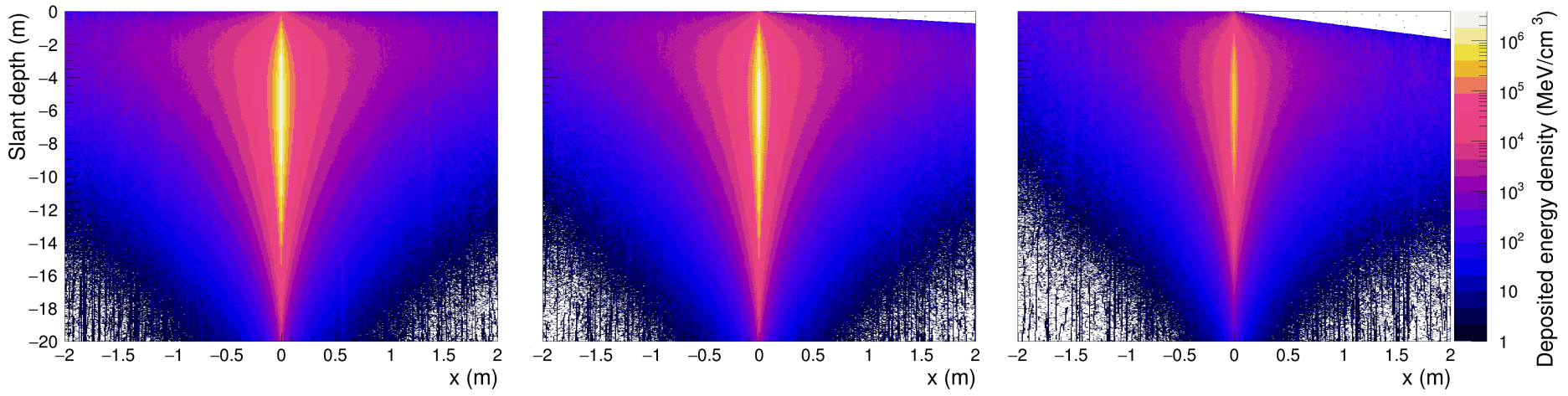}
		\caption{Deposited energy densities for $E_p = \SI{e18}{eV}$. From left to right, the zenith angle values are $\theta = 0^{\circ}$, $\theta = 20^{\circ}$ and $\theta = 40^{\circ}$. }
		\label{fig:slice_histos_1e9} 
	\end{subfigure}
	\caption{The deposited energy density within a vertical 1-cm wide slice going through the center of the particle shower, for the reference air shower at different primary energies $E_p$ and zenith angles $\theta$. The y-axis shows slant depth, here defined as the depth in unit length along the shower axis with respect to the ice surface. In this reference system the ice surface is tilted clockwise over an angle $\theta$, showing up in the upper right corners of the plots. The color scale is fixed for each value of $E_p$ and shown on the right of the plots.}
	\label{fig:slice_histos_appendix}
\end{figure*}

\clearpage

\section{Fitted parameters to the average $w_1(r)$ distributions}\label{sec:AppB}

\def\arraystretch{1.5}
\setlength{\arrayrulewidth}{0.5mm}
\begin{table}[h!]
	\begin{tabular}{ |c|c|c| } 
		\hline
		$X_{max}$ (\SI{}{\g/\cm\squared}) & $b$ (\SI{}{\m}) & $c$ \\
		\hline
		538 - 652 & $(7.9 \pm 1.2) \times 10^{-2}$ & $0.774 \pm 0.081$ \\ 
		652 - 766 & $(3.50 \pm 0.27) \times 10^{-2}$ & $0.645 \pm 0.020$\\ 
		766 - 880 & $(2.38 \pm 0.18) \times 10^{-2}$ & $0.585 \pm 0.014$ \\ 
		880 - 994 & $(2.33 \pm 0.17) \times 10^{-2}$ & $0.617 \pm 0.015$ \\ 
		994 - 1109 & $(9.31 \pm 0.50) \times 10^{-3}$ & $0.5161 \pm 0.0068$ \\ 
		\hline
	\end{tabular}
	\caption{\label{tab:fit_vals} The fitted values of the parameters of Equation~\ref{eq:fit_function} for the different average $w_{1}$ distributions at a cascade front depth of \SI{450}{\g/\cm^2} with respect to the ice surface.}
\end{table}

\end{document}